\documentclass[twocolumn]{aastex62}
\usepackage{amsmath}
\usepackage{graphicx}
\usepackage{natbib}

\received{February 10, 2019}
\submitjournal{ApJ}

\shorttitle{Machine learning on planetary collisions}
\shortauthors{Valencia et al.}

\begin{document}

\title{Can a machine learn the outcome of planetary collisions?}

\correspondingauthor{Diana Valencia}
\email{valencia@astro.utoronto.ca}

\author{Diana Valencia}
\affil{Centre for Planetary Sciences \\
University of Toronto \\
1265 Military Trail \\
Toronto, ON, M1C 1A4, Canada}

\author{Emaad Paracha}
\affil{Centre for Planetary Sciences \\
University of Toronto \\
1265 Military Trail \\
Toronto, ON, M1C 1A4, Canada}

\author[0000-0003-4393-9520]{Alan P. Jackson}
\affil{Centre for Planetary Sciences \\
University of Toronto \\
1265 Military Trail \\
Toronto, ON, M1C 1A4, Canada}

\begin{abstract}
Planetary-scale collisions are common during the last stages of formation of solid planets, including the Solar system terrestrial planets. The problem of growing planets has been divided into studying the gravitational interaction of embryos relevant in million year timescales and treated with $N$-body codes and the collision between objects with a timescale of hours to days and treated with smoothed-particle hydrodynamics. These are now being coupled with simple parameterized models. We set out to investigate if machine learning techniques can offer a better solution by predicting the outcome of collisions which can then be used in $N$-body simulations. We considered three different supervised machine learning approaches: gradient boosting regression trees, nested models, and gaussian processes. We found that the former produced the best results, and that it was slightly surpassed by ensembling different algorithms. With gaussian processes, we found the regions of parameter space that may yield the most information to machine learning algorithms.  Thus, we suggest SPH calculations to focus first on mass ratios above 0.5.
\end{abstract}


\section{Introduction}
\label{sec:intro}

A widely accepted prediction of current planet formation theories is that the final stage in the growth of solid planets like our own terrestrial planets is one of chaotic growth via giant, planetary-scale impacts between embryos \citep[e.g.][]{kenyon2006, raymond2009, kokubo2010}.  These collisions are dramatic events and can lead to substantial changes in the bodies involved, most obviously in mass, but also potentially in composition or rotational state \citep[e.g.][]{asphaug2010, marcus2010a, marcus2010b, movshovitz2016}.  Despite the wide array of outcomes that giant impacts can display, from efficient accretion to catastrophic disruption and hit-and-run collisions, modelling of the formation of systems of solid planets has typically assumed that collisions always result in perfect mergers with no debris\citep[e.g.][]{Chambers:2001,Kokubo:2002,Raymond:2004,Terquem:2007,Bond:2010}.  This has largely been a result of computational limitations, since computational time increases rapidly with the number of particles in an $N$-body code, even when employing algorthmic tricks to reduce the number of force calculations.

As computer power has increased authors have begun efforts to incorporate more realistic collision treatments into models of planet formation.  The first such successful effort was that of \citet{kokubo2010} who incorporated a simple switch allowing a collision to be either a perfect merger or a hit-and-run collision in which both target and impactor were unchanged depending on the impact velocity and impact angle.  Since then more sophisticated treatments have been constructed, including a limited number of debris particles \citep[e.g.][]{Chambers:2013, Mustill:2018}.

Behind all of these efforts at incorporating giant impact outcomes into $N$-body simulations is a model for predicting the outcome of an impact based on the impact parameters taken from the $N$-body integration.  Giant impacts unfold over timescales of hours rather than the millions of years with which $N$-body planet formation simulations are concerned.  Although the planets we are interested in are composed of rocky materials, the dynamics of the collision between bodies large enough to be in hydrostatic equilibrium are overwhelmingly dominated by gravity such that it can be treated as a fluid dynamics problem using standard fluid dynamics techniques, with the most widely used being Smoothed-Particle Hydrodynamics (SPH).  There have been efforts at directly combining $N$-body and SPH such that whenever a collision occurs in the $N$-body integration an SPH simulation of the collision is spawned \citep[e.g.][]{genda2011}, but despite increases in computing power this remains unfeasible.  As such, a simplified model mapping the space of impact parameters onto the space of impact outcomes is needed.

The accuracy of new generation $N$-body models that incorporate more realistic giant impact outcome treatments is thus dependent on how accurate the impact models are.  The potential input parameter space for giant impacts is large; the mass ratio, impact velocity and impact angle are commonly considered to be the three most important parameters but the outcome can also be influenced by the absolute mass, material composition, thermal state, material strength, and pre-impact rotational state of the two bodies, with the latter itself requiring 6 independent variables.  Giant impact simulations themselves are also still quite expensive, a typical SPH simulation at reasonable resolution takes at least several days even on modern hardware.  As such, efforts at mapping the space of giant impact input variables onto the space of outputs have focussed on only a subset of the potential input variables.  For example, the most widely used such model at present is that of \citet{Leinhardt_Stewart:2012} or LS12 hereafter, but this is only based on around 200 simulations with limited sampling in absolute mass, mass ratio, impact velocity and impact angle.

In this study, we set out to investigate whether or not machine learning can predict the outcomes of planetary collisions with the goal of enabling the study of planet formation with more realistic collisional outcomes. 

In the following sections we will describe the data set we used to train, validate and test our machine learning model (section 2), give a brief introduction to machine learning and describe our model techniques (section 3), followed by our results (section 4) and summary (section 5).

\section{Data}
\label{sec:data}

\begin{figure}
\begin{centering}
\includegraphics[width=\columnwidth]{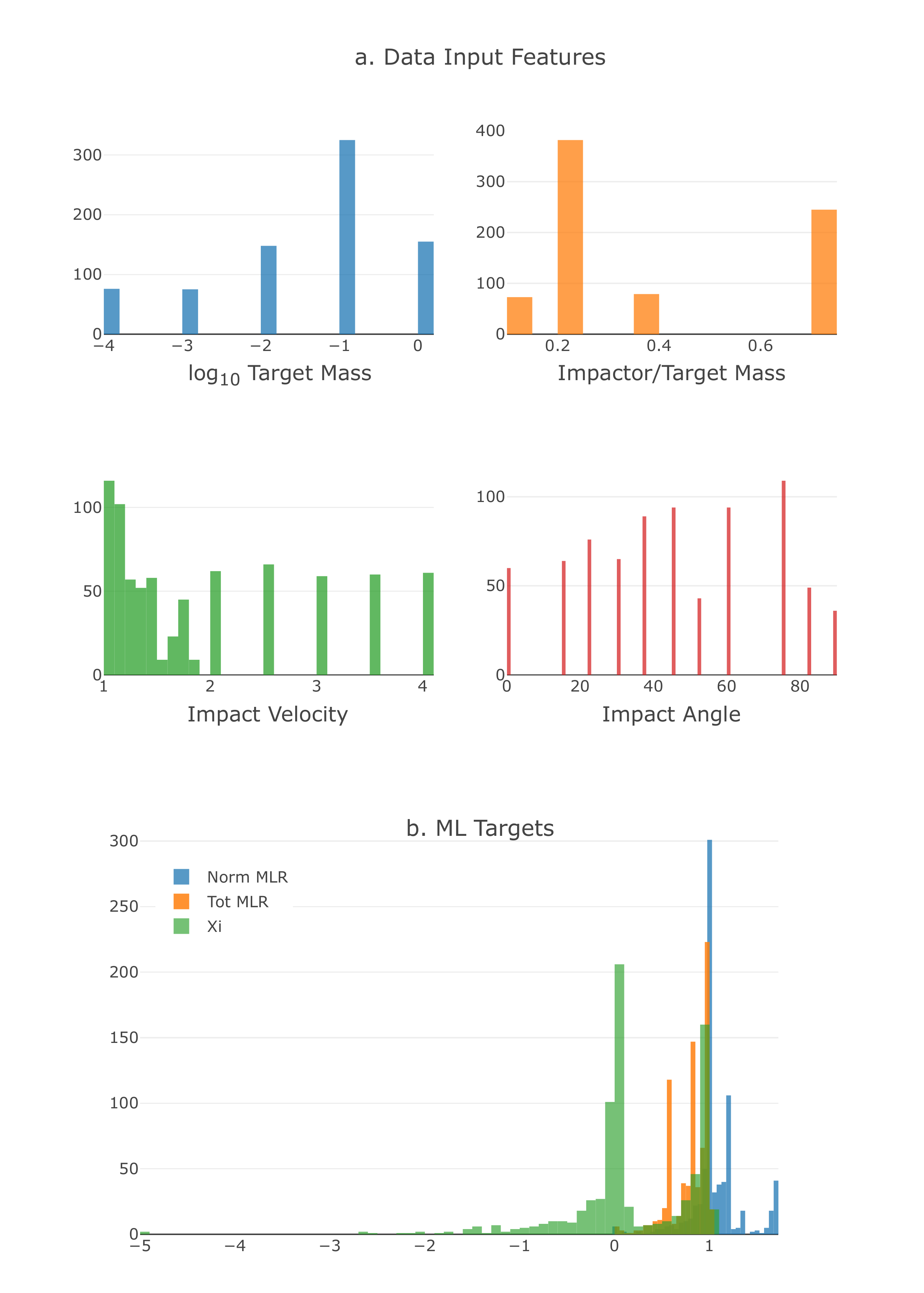}
\end{centering}
\caption{Histogram of collisional data. a) Input features, and b) Machine learning targets: $M_{\rm LR}/M_{\rm tar}$ (blue), $M_{\rm LR}/M_{\rm tot}$ (orange), $\xi=\left(M_{\rm LR}-M_{\rm tar}\right)/M_{\rm imp}$
(green) \label{fig:Histogram-data}}
\end{figure}

For our ground truth on giant impact outcomes we use the data provided by Gabriel et al. (2019, submitted), hereafter G19, specifically their differentiated rock-iron database, which is the most well sampled of the three compositions they consider.  This database consists of 1039 SPH simulations of collisions between bodies composed of 30 wt\% iron and 70 wt\% quartz (SiO$_2$), with both materials modelled using the ANEOS equation of state.  Quartz was chosen to represent the rock component since it has the most up to date equation of state \citep{melosh2007} and has been widely used in previous studies \citep[e.g.][]{canup2001, marcus2009, asphaug2014}.  Each simulation was conducted with $10^5$ particles in the target body.  The input parameters that are varied within the database are the mass of the target ($M_{\rm tar}$, ranging from 0.01 to 1 $M_{\oplus}$), the ratio of impactor to target masses ($\gamma=M_{\rm imp}/M_{\rm tar}$, ranging from 0.1 to 0.7), the impact velocity normalized to the escape velocity ($v_{\rm imp}/v_{\rm esc}$, ranging from 1 to 4) and impact angle ($\theta$, ranging from 0$^{\circ}$ to 89.5$^{\circ}$).  A histogram of these parameters is shown in Fig.~\ref{fig:Histogram-data}.  The most well sampled parameters are the impact velocity and the impact angle, with the target mass and impactor to target mass ratio more sparsely sampled.  For more information on the construction of the database and the rationale for the ranges of parameters chosen, please refer to G19.

The output of an SPH simulation contains a large number of properties of potential interest, including the mass distribution of the bodies, their velocities, temperatures and spins.  For the purpose of this study we are focussed on the mass of the largest remnant ($M_{\rm LR}$), or conversely the accretion efficiency parameter $\xi=\left(M_{\rm LR}-M_{\rm tar}\right)/M_{\rm imp}$ defined by \citet{Agnor_Asphaug:2004} which measures the change in the mass of the target in units of the mass of the impactor.   A value of $\xi=1$ corresponds to perfect accretion, a value of  $\xi \sim 0$ corresponds to a hit-and-run collision where the mass of the target and impactor are largely unchanged, and a value of $\xi<0$ signifies erosion of the target.

In the context of machine learning problems, 1039 samples may be considered small. However, the minimum size of a dataset that is needed for useful predictions depends both on how complex the system one is trying to model is, and the purpose of the predictions.  The mass of the largest remnant can be considered in many ways the most basic, first order outcome of a giant impact, hence our focus on this quantity in investigating if machine learning can be a useful tool for predicting collision outcomes for $N$-body formation models.

\section{Machine learning methodology}
\label{sec:ml}

Given the size of the data set, we concentrated on machine learning techniques that are known to perform well in small datasets (gradient boosting trees and gaussian processes). We use these in three different approaches: gradient boosting trees for regrssion, gaussian processes and nested models (classification and regression). A brief heuristic description of supervised machine learning and each of the models used here follows.

Supervised machine learning aims to predict an outcome or target given some data features, by training an algorithm on previous data where all the 
variables are known. The goal is to have a trained algorithm that has learned the `rules' of the system with the training data, which can then be applied to future, new data. In supervised machine learning, training is achieved by minimizing a loss function (e.g. sum of the errors) and confidence in the model is often based on the algorithm achieving some desired `accuracy' or other meaningful metric on test (unseen) data. A common practice is to split the data into training, validation and test data. The training set is used to train the model, the validation set is used to optimize the algorithm hyper-parameters, and the test data is used to verify that there is no data leakage (or overfitting).  To maximize the use of data, we used cross-validation (CV) on the
training-validation set, where the data is divided into $n$ folds, or groups, and each fold acts as a validation set once, with the rest of the folds being used for training, yielding $n$ models that are then averaged together. Supervised machine learning problems can be divided into regression problems, where the goal is to predict a number (eg. the selling price of a house), or classification problems, where the goal is to predict the class a particular input belongs to (e.g. is it a cat or a dog?).

It is very important to note that the cross-validation approach is fundamentally different than fitting techniques that use all the data available to fit parameters in a proposed model, which is common practice in physical systems. For example, the efforts from LS12, \citet{movshovitz2016}, and G19, all use the latter approach. They propose a model inspired by our physical understanding of planetary collisions to suggest a set of equations with few parameters that they then fit to using the whole data set. Instead, we only train the ML algorithm using  part of the data, and choose the best hyper-parameters by comparing performance in the validation set. By regularizing, we make sure we penalize overly complicated models that would perform very well on known data, but expected to fail when fed new unseen data.  Lastly we feed the model the test (unseen) data set to ensure there is no information leakage. In contrast, our approach emulates the capacity of the model to predict on new data.

\subsection{Models}
\label{sec:ml:models}

For this study we used two very different machine learning techniques, gradient boosting trees and gaussian processes, with three different approaches. We applied:

\begin{enumerate}
 \item straight gradient boosting trees as a regression problem to the machine learning target (ml-target),
 \item gaussian processes as a regression problem and obtained error bars associated with the prediction, and 
 \item a nested approach to first classify the outcome (using gradient boosting classification trees) into four different collisional regimes, and then apply a regression scheme to each of the classes. The four categories for the classification were:
\begin{enumerate}
 \item perfect accretion: when the impactor and target merge inelastically, equivalent to an efficiency of $\xi=1$.
 \item partial accretion: when the mass of the largest remnant is larger than the target's mass $\xi>0$. 
 \item hit and run: when the impactor nearly misses the target leaving the target nearly unchanged, equivalent to an efficiency of $\xi=0$
 \item partial erosion: when the target loses part of its mass $\xi<0$.
\end{enumerate}

\end{enumerate}

In practical terms, we explored different boundaries for the different classes and settled on perfect accretion defined as $\xi > 0.95,$ partial accretion as $0.95 \ge \xi > 0.1,$ hit and run as $0.1 \ge \xi \ge -0.1$ and partial erosion as $\xi<-0.1$.

In machine learning, the choice for the loss function is an integral part of producing a useful model. A loss function defines the metric by which to measure the performance of the model. For example, while both the mean absolute error and the mean squared error (MSE) express averages of the deviation between the predicted outcome and ground truth, because the errors are squared before they are averaged with MSE, this metric gives more weight to outliers. This is a useful property because we consider it is best  to minimize large errors in the prediction of the mass of the largest remnant for the goal of implementing planetary collisions into $N$-body codes. In other words, it is better to have medium errors an all collisional predictions than to have small errors on most, but large errors on a few intractable predictions.  Thus, we opted for MSE as our choice for loss function. However, we also investigated the mean absolute error and different custom errors, finding that they did not produce much overall change in the models over using MSE. For all models we used a 5-fold cross-validation, as the dataset is small and we aimed to minimize stochastic variability in MSE among folds.

\subsection{Gradient Boosting Regression Trees}
\label{sec:ml:grboost}

A common algorithm in machine learning is to use decision trees, which split the data into a tree-like structure (if/else statements or rules) using the values of the independent variables to get to the possible outcomes or values of the dependent variables. These trees can be simple and have few branches, or have a more complicated structure with many branches, as well as different numbers of decision levels (depth of the trees). Gradient boosting involves using the residuals of this model and fitting them to a subsequent model, in an attempt to correct the errors of the previous modelling stage. The addition of a new model of the residuals is applied many times so as to focus on the `hard to predict points' that will dominate the residuals at each iteration. This is done until the model starts overfitting (modeling the noise) or the sum of the residuals does not improve. 

In practical terms, we use the $\mathtt{XGBoost}$ \citep{Xgboost} open-source algorithm as integrated in the $\mathtt{scikit-learn}$ package in python, and a grid search approach on the number of estimators (or trees) and the maximum depth.

\subsection{Gaussian Processes}
\label{sec:ml:gp}
Gaussian processes (GP) are a non-parametric approach to machine learning that finds a distribution over a possible set of functions $f(x)$ that are consistent with the observed data. This stands in contrast to the bayesian approach that finds a distribution over the parameters of a chosen model. A gaussian process assumes that the joint probability of the possible functions $p(f(\mathbf{x}_{1}),...,f(\mathbf{x}_{N}))$ is jointly gaussian, with some mean $\mu(\mathbf{x})$ and covariance $\Sigma(\mathbf{x})$ given by a positive definite kernel function $\Sigma_{ij}=k(\mathbf{x}_{i},\mathbf{x}_{j})$. If two points are considered by the kernel to be close in space, the output of the function at those points is similar too. By looking at the distribution of functions that can fit the observed data, GP yields an error associated with the prediction, which is small at the points where there is data (with errors), large between data points, and larger at points where extrapolation is required. We used the $\mathtt{george}$ package by \citet{george} for python.

Our kernel of choice was the squared exponential in four dimensions (corresponding to the four simple features of the data). We considered a linear combination of square exponentials for which we obtained the hyper-parameters of the amplitude and characteristic length-scale of each, via an optimization scheme (using the open source package $\mathtt{hyperopt}$ \citet{hyperopt}):

\[
k(\mathbf{x}_{i,}\mathbf{x}_{j})=\sigma_{1}^{2}\exp\left(\frac{\mathbf{x}_{i}-\mathbf{x}_{j}}{2\ell_{1}}\right)^{2}+\sigma_{2}^{2}\exp\left(\frac{\mathbf{x}_{i}-\mathbf{x}_{j}}{2\ell_{2}}\right)^{2},
\]

where $\sigma_{1},\sigma_{2},\ell_{1},\ell_{2}$ are the hyper-parameters we optimized for. Having two squared exponential functions allowed us to better capture small and large scale variations in the predictions compared to one square exponential only. We also tested three but there was no gain. As a first study, we did not further test other types of kernels and instead leave this for future work. More information on gaussian processes can be found in \citet{GPbook}.

\subsection{Feature Engineering}
\label{sec:ml:features}

We first considered four basic features taken directly from the SPH simulation inputs: target mass, ratio of impactor mass to target mass, ratio of impact velocity to escape velocity, and impact angle.  We also tried a variety of combinations of features from LS12, including the ratio of the impact energy to the critical impact energy needed to disrupt the body (inspired by their dominant term in their proposed universal law), or ratio of impact parameter to the critical impact parameter that is used to differentiate between grazing and non-grazing collisions, or the universal law (equation 5 of LS12) applied indistinctively to all samples, and many others in an effort to provide physical quantities to the model. Unfortunately,  these features were not useful to the model. 

Subsequently, we used the prescription provided by LS12 in the appendix to obtain their suggested $M_\mathrm{LR}/M_\mathrm{tot}$ which we converted to a collisional efficiency $\xi^{LS}$.   This prescription uses switches to differentiate between the different regimes: perfect accretion, partial accretion, hit and run, partial erosion or catastrophic, and super-catastrophic regimes. They suggest the outcome of perfect accretion to be set as $ M_\mathrm{LR}=M_\mathrm{tot}$,  partial accretion to be calculated from their proposed universal law, hit and run to be set as $ M_\mathrm{LR}=M_\mathrm{tar}$, catastrophic regime to follow their universal law, and super-catastrophic to follow a modified universal law (Eq. 44 in their paper). There are only two parameters ($c*$ and $\bar{\mu}$ ) required for their universal law and whose values were obtained by LS12 via a fit to their data . Given that our dataset comprises collisions between large bodies, we used their corresponding suggested values of $c*=1.9$ and $\bar{\mu}=0.36$.   Another feature we used was their collisional categories.

Likewise, we also used the collisional prescription suggested by G19. They propose a relation for $M_\mathrm{LR}/M_\mathrm{tot}$ based on an overall fit to the same dataset we are using. The only switch they use is one that identifies if a collision is in the hit-and-run regime or not. We used both a calculated efficiency $\xi^{G}$ from their work and a collision type (as being in the hit-and-run regime or not) as our engineered features.

Lastly, we considered geometric features $\left(\frac{v_{\rm imp}}{v_\mathrm{esc}}\right)^2$, 
 $\theta^2$, 
 $\frac{1}{\theta} \times \left(\frac{v_{\rm imp}}{v_\mathrm{esc}}\right)$, 
$ \textcolor{red}{[??]}\times \left(\frac{v_{\rm imp}}{v_\mathrm{esc}}\right)^{2}$, 
$ \frac{1}{\theta^2}  \times \left(\frac{v_{\rm imp}}{v_\mathrm{esc}}\right)$, 
$  \frac{1}{\theta^2} \times \left(\frac{v_{\rm imp}}{v_\mathrm{esc}}\right)^{2}$,
 $\theta \times \left(\frac{v_{\rm imp}}{v_\mathrm{esc}}\right)$, 
 $\theta^{2} \times \left(\frac{v_{\rm imp}}{v_\mathrm{esc}}\right)$, 
 $\theta \times \left(\frac{v_{\rm imp}}{v_\mathrm{esc}}\right)^{2}$ 
 in the hopes of helping distinguish the regimes of high erosion (typically found at low impact, high velocity), hit and run (high impact angle), and perfect merger (low impact velocity, low impact angle).

We found very marginal improvements with all these engineered features. Despite the tens of models we built with engineered features we were not successful at significantly improving the predictions over using only the four basic features. See more below.

\section{Results}
\label{sec:results}

The first task was to decide on a good machine-learning target. Given that the largest remnant masses in the data span several orders of magnitude, stemming from orders of magnitude variation in the target mass, it is necessary to normalize $M_{\rm LR}$ to avoid spurious correlations. There are at least three possibilities for normalization: (i) the mass of the largest remnant normalized to the target mass $M_{\rm LR}/M_{\rm tar}$, (ii) the mass of the largest remnant normalized to the total mass $M_{\rm LR}/\left(M_{\rm tar}+M_{\rm imp}\right)=M_{\rm LR}/M_{\rm tot}$, and (iii) the accretion efficiency $\xi=\frac{M_{\rm LR}-M_{\rm tar}}{M_{\rm imp}}$. From the distributions (see Figure~\ref{fig:Histogram-data}) we see that there are a few peaks in the normalized $M_{\rm LR}$ (to target mass) values, owing to clusters around hit and run ($M_{\rm LR}/M_{\rm tar}=1$), and perfect accretion of the sampled mass ratios $0.1,\,0.2,\,0.3,\,0.7$. The distribution of $M_{\rm LR}/M_{\rm tot}$ has a similar behaviour although less tractable. On the other hand, the distribution of the efficiency has two very distinct peaks for hit and run ($\xi=0$) and perfect accretion ($\xi=1$).

We tested all three ml-targets by comparing the mean squared errors on predicted $\xi$ for the test set. The values were 0.054 for $M_{\rm LR}/M_{\rm tar}$, 0.053 for $M_{\rm LR}/M_{\rm tot}$, and 0.041 for $\xi$. We opted to train on $\xi$. 
However, MSE is a metric that is limited in its use, as it only provides an average error. Some predictions can be substantially wrong and we would not know which ones they are.  Thus, as another way to judge the goodness of the prediction, we calculated how many samples were well-predicted and within a 0.25 margin, versus how many were beyond.   That is, samples that fell within $-0.25 \leq \Delta \xi \leq 0.25$, where $\Delta \xi$ is the difference between the ground truth and the predicted collisional efficiency, were considered well-predicted. This metric showed values of 86\%, 89\% and 88\% for the test samples, respectively.  We note that our choice of setting a margin around $\xi$ instead of $M_{\rm LR}/M_{\rm tot}$ like in LS12 or G19, is more stringent as the denominator of $\xi$, $M_{\rm imp}$, is smaller than $M_{\rm tot}$ and amplifies any errors.  These metrics and those for all models explored are shown in Table \ref{performance}. 

\begin{table*}
\centering
\begin{tabular}{ c | c c c c c c} \hline 
Model & ml-target  & MSE validation & MSE test & Test set  & Overall set   & Overall set   \\ 
          &                  &                          &                &  0.25 margin      &            0.25 margin        &    0.10 margin \\ \hline  
GBRT basic features  &  $\xi$                                               & 0.0312$\pm$0.007 & 0.041 & 88\% & 91\% & 77\%  \\    
GBRT basic features & $M_\mathrm{LR}/ M_\mathrm{tar}$ & 0.038                  & 0.05 & 86\% & 91\% & 74\%  \\ 
GBRT basic features & $M_\mathrm{LR}/ M_\mathrm{tot}$ & 0.025                  & 0.05 & 89\% & 91\% & 72\%  \\  \hline

GBRT basic features + $\xi^{G}$  &  $\xi$                              & 0.031$\pm$0.006  & 0.042 & 90\% &  90\% & 76\%  \\ 
GBRT basic features + $\xi^{LS}$ &  $\xi$                             & 0.033$\pm$0.008  & 0.045 & 85\% 90\% & 77\%  \\ 
GBRT Trees - basic and geometric features & $\xi$              & 0.026$\pm$0.006 & 0.041 & 90\% & 92\% & 78\%  \\   \hline

Nested - basic features & $\xi$                                               & 0.043                & 0.051 & 91\% & 90\% & 75\%   \\
Nested - basic features and $\xi^{LS}$ & $\xi$                       & 0.043                & 0.043 & 89\%  & 0.90\% & 73\%   \\
Nested - basic features and geometric features & $\xi$         & 0.044                & 0.045 & 90\%  & 0.90\% & 71\%   \\

Gaussian Processes basic features  & $\xi$                           & 0.045$\pm$0.006 & 0.057 & 76\% &  89\% & 51\%  \\ \hline


\textbf{Ensembled} & $\xi$                                                                  &  0.025                 & 0.037 & 88\% & 92\%   & 74\% \\ \hline

LS12  &                                                                                   &  0.15*                &   & -  & 73\%   & 55\% \\ 
G19 &                                                                                      &  0.14*                 &   & - & 74\%   & 45\% \\ \hline

\end{tabular} 
\caption{Performance of the Different Machine Learning Models \protect \\
\hspace{1em}\tiny *MSE of all samples 
\normalsize 
}
\end{table*}
\label{performance} 

In addition, we found ten problematic samples that had a ground-truth value for the largest remnant mass to be zero. Given that this value is not physical, as there should always be some debris left even after super-catastrophic collisions, we followed G19 and removed them, attributing them to an error in the SPH identification. Removing these samples considerably improved the MSE of the predictions. Also, we found 29 samples with an efficiency slightly greater than one, which we modified and set to 1 in accordance to what is physically possible. This latter practice yielded slight gains in the MSE of the predictions.

\subsection{Gradient Boosting Trees Regression}
\label{sec:results:grboost}

\begin{figure}
\begin{centering}
\includegraphics[width=\columnwidth]{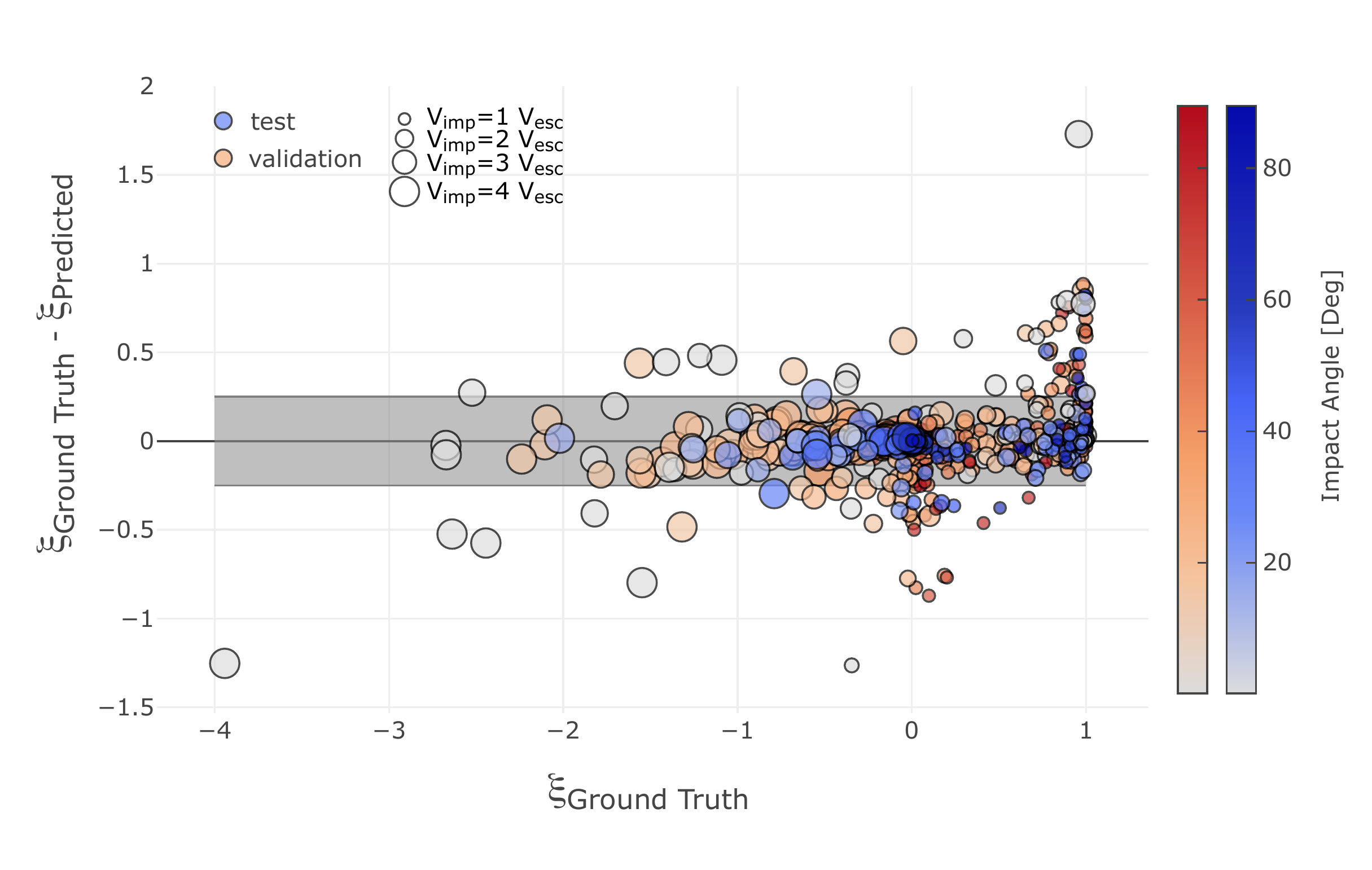}
\end{centering}

\caption{Differences between Ground Truth and Predicted values for Collisional Efficiency using Gradient Boosting Trees Regression. Prediction of out-of-fold cross validated data (red) and test data (blue) are colour-coded according to impact angle, and sized according to the ratio of impact velocity to escape velocity. Only the four basic features were used (target mass, mass ratios, impact velocity to escape velocity, and impact angle). The shaded region indicates the 0.25 margin perfect predictions. \label{fig:Comp-GBTR-4F}}

\end{figure}

We split the data into training and test sets in a 4:1 ratio, and trained a gradient boosting regression trees (GBRT) model using cross-validation on the train set first with only the four simple features.
We used a grid search approach with cross-validation to find the best hyper-parameters. With these optimal hyper-parameters under cross-validation we obtained predictions on the out-of-fold validation data, which we then compared to the ground truth within the training set. We subsequently retrained the model with the optimal hyper-parameters and the whole training data set, and used it to predict on the test set. This test set is mostly used to make sure we are not overfitting or leaking data into the predictions. Figure \ref{fig:Comp-GBTR-4F} shows the comparison between the ground-truth and the predictions of the efficiency for the out-of-fold validation (red) and for the test data set (blue). It is shaded according to impact angle, and sized according to $v_{\rm imp}/v_{\rm esc}$.  The mean square error for the out-of-fold validation samples is $0.0312\pm 0.0074$ and for the test set is $0.041$, consistent at the 1.5-$\sigma$ level. We tested several random seed values used in selecting the test and validation set and obtained consistent results. That is, the variations in the MSE of the test set arising from different random seeds were consistent with the MSE variations within each validation set, suggesting minimal overfitting.  

\begin{figure}
\begin{centering}
\includegraphics[width=\columnwidth]{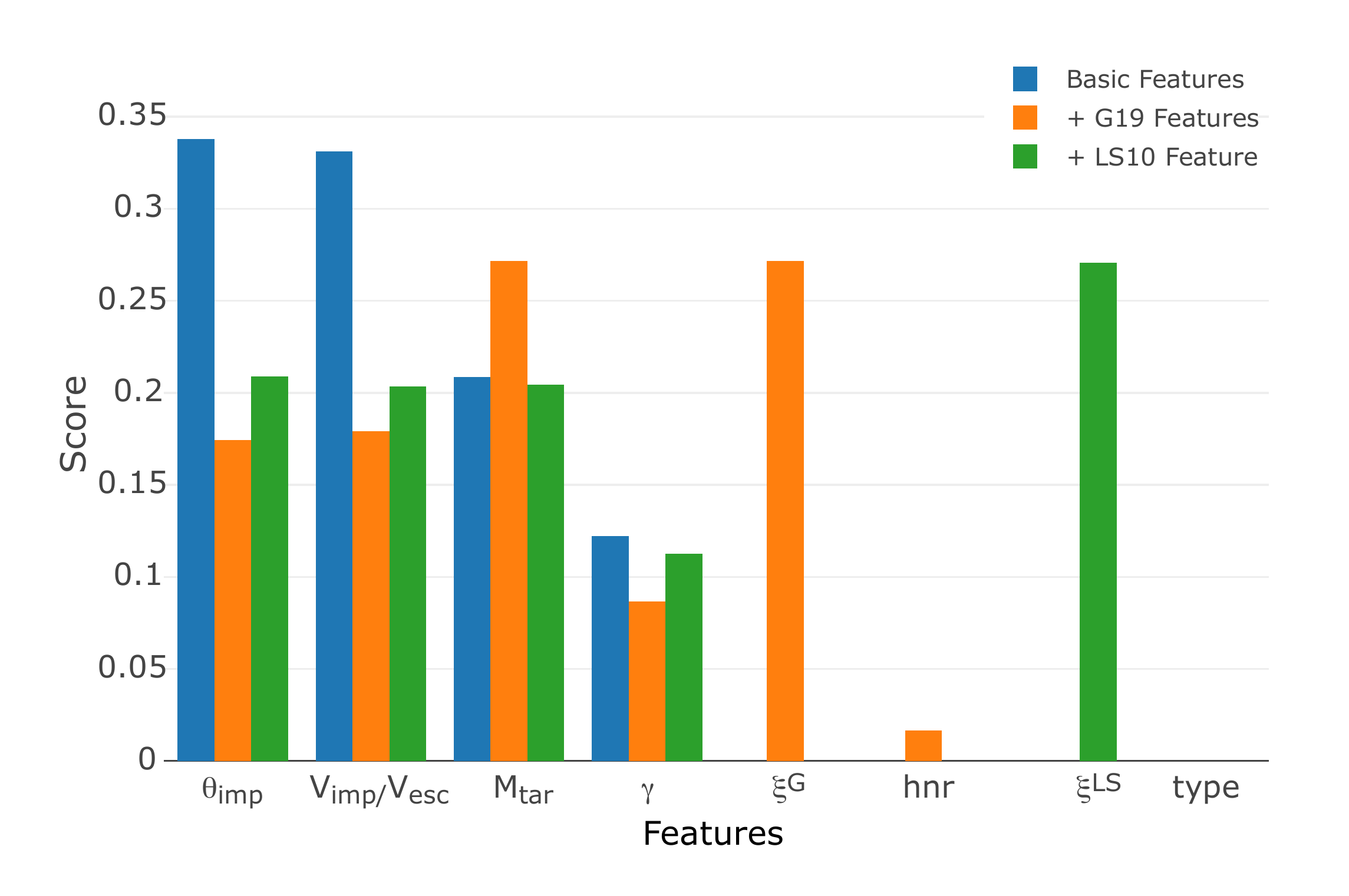}
\end{centering}

\caption{Feature Importance for models constructed using 1) the four simple features (blue), 2) adding the efficiency $xi^{LS}$ and type of collision (type$^{LS}$) from LS12 (orange), 3) adding instead the efficiency $xi^{G}$ and a switch to identify hit-and-run (hnr$^{G}$) from G19 (green), and 4) adding instead the geometrical features (red) . Y axis shows the relative score of each feature in contributing to the decision trees model, normalized so that all scores add up to one.  \label{fig:FI}}
\end{figure}

We found that all the four basic features were used by the model for predicting the outcome (see Fig. \ref{fig:FI}). This model yielded 88\% of the test samples within the 0.25 error margin and 81\% within the 0.1 error margin.  However, because of low number statistics (there are only 206 test samples) we also show in Table 1 the percentage of well-predicted cases for all samples (via the validation and test sets) and annotate that these are most likely optimistic values given the model is tuned to predict the validation set.  We note that there is a trend for the model to under-predict $\xi$ when the ground truth value is close to 1 (i.e. perfect accretion) and to over-predict when the ground truth is close to 0 (hit-and-run). This is at least in part due to values of $\xi \ge 1$ being unphysical, and as shown later the same pattern is evident in the models of LS12 and G19.

We then used the feature-engineering approach. We used the two added features from LS12 which where the calculated efficiency predicted from their prescription, and a classification into five different collisional regimes (perfect accretion, partial accretion, hit-and-run, partial erosion, super-catastrophic). We saw no improvements in the model, and a slight worsening of the  MSE score on unseen data (0.045).  However, the important features for the model changed, with their estimated efficiency $\xi^{LS}$ being the most important contributor (see Fig. \ref{fig:FI}).

We subsequently used the features inspired by the prescription from G19, which also included an estimated efficiency, and a classification either into hit-and-run or not. This yielded similar scores as our nominal model.  Like with the features from LS12,   their estimated efficiency $\xi^G$ became the most important feature in predicting the outcome (see Fig. \ref{fig:FI}).   When we tried different random seeds we saw variation in the importance of $\xi^{G}$ and $\xi^{LS}$ for the predictions. However, for all random seeds our nominal model performed well with the four basic features alone, indicating these are enough for the algorithm to learn about the collisional outcomes.  

Finally, we tried adding the geometrical features to the four basic features and obtained an improvement in the MSE score of the validation set  but no significant improvement on the test set. Figure \ref{fig:FI} shows the importance of each feature in determining the results for these three models.  Table 1 shows the metrics. In some cases better MSE scores reflect an improvement on the poorly-predicted samples but not enough to be considered within the 0.25 margin.

\begin{figure}
\begin{centering}
\includegraphics[width=\columnwidth]{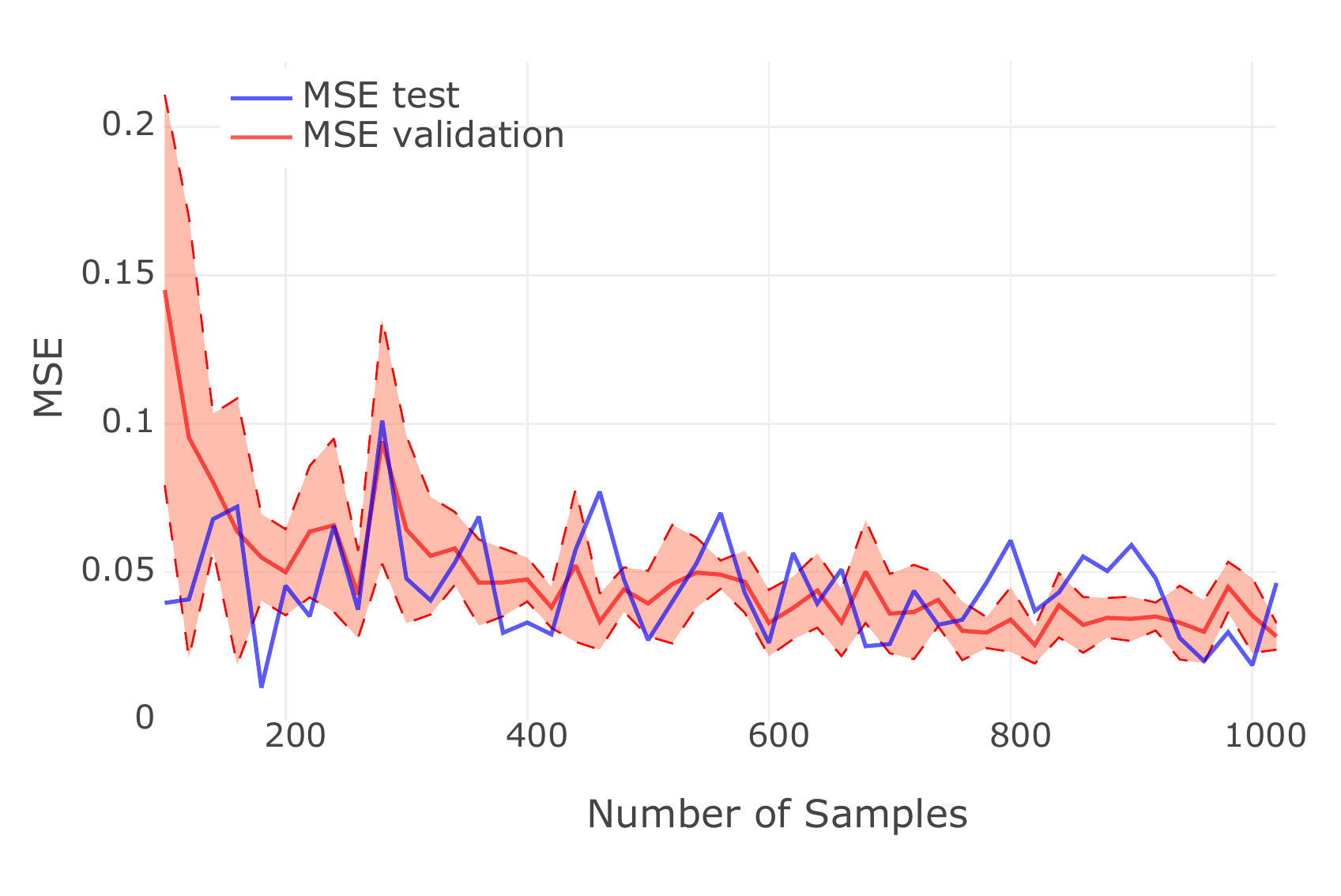}
\end{centering}

\caption{Learning Curve for prediction on Collisional Efficiency using Gradient Boosting Trees Regression. The MSE error in the validations set (red) and test set (blue) are shown. \label{fig:learning}}
\end{figure}

The last test we performed to investigate the prediction power of machine learning on this small data set was to hide part of the data and ask by how much the predictions worsened. We took subsets of the data randomly with samples ranging from only 100 to the full data set. For each subset we optimized for hyper-parameters, calculated the MSE on the validation set, and the test data.  Figure~\ref{fig:learning} shows the results. We find that adding more data improves the MSE of the validation set up and that the size of the data is beginning to reach a floor value. In contrast, the MSE of the test data does not seem to decrease with more samples nor its variability. Thus, it seems that the pathway to improve the prediction power of machine learning algorithms consists of sampling a very different parameter space in the SPH simulations and not just to acquire more data probing the existing space.

\subsection{Nested Models}

Motivated by the fact that there is a spread in the predictions at ground-truth values of $\xi=0$ and $\xi=1$ (see Fig. \ref{fig:Comp-GBTR-4F}) , which correspond to hit-and-run and perfect accretion regimes, we decided to use a nested model approach in which we first classify the collision output into perfect accretion, partial accretion, hit-and-run and partial erosion before using a regression model within each category. The idea is to test whether sub models can perform better. In general, one expects the more data seen by the algorithm the better it performs, however, because the regressor is having difficulty with particular types of collisions it is worth exploring building sub-models.

\begin{figure}
\begin{centering}
\includegraphics[width=\columnwidth]{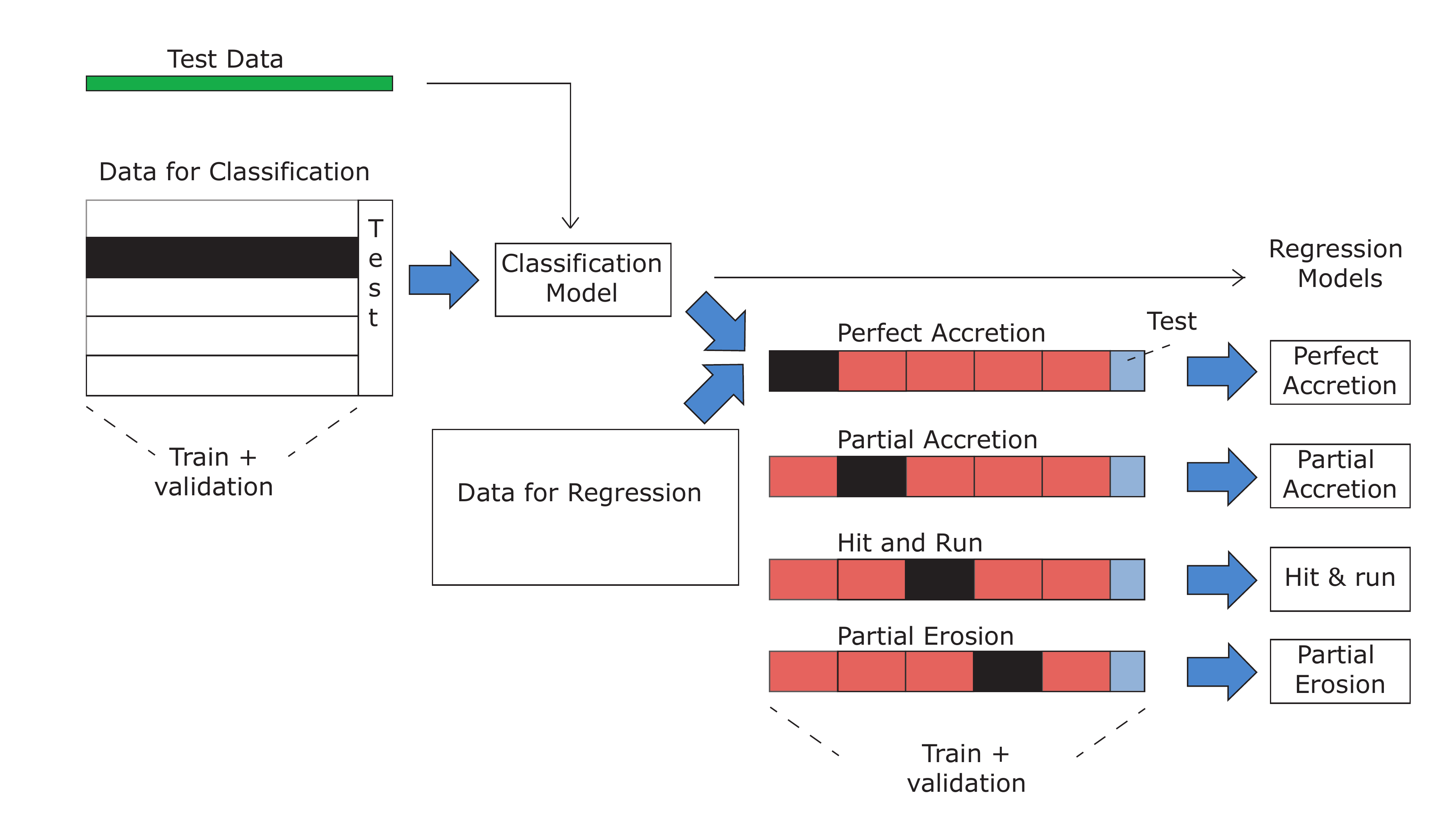}
\end{centering}

\caption{ Schematic diagram of how the data was split and used for the nested models. The data was split 10\% for overall testing, 45\% for building the classification model and 45\% to build each regression model for each type of collision once the data has been classified by the classification model.  For each model, we used a cross-validation with four folds. \label{fig:schematic}}

\end{figure}

We took a test split of 10\%, and the rest was split in half for the classification and regression models. For the classification part, out of this half, 10\% was set for testing, and 20\% for validation under CV. Then the data that was set aside for regression was fed into the classifier (as new data) and each of the classes was split into 10\% test, and 20\% for validation under CV. Finally the original test sample was fed into the classifier and then the regressor as unseen data for the nested model approach to be tested. A schematic diagram of how we handled the data is shown in Fig. \ref{fig:schematic}.

Relevant classification measures are shown in Table \ref{nested}, where precision captures the fraction of false-positives, recall the fraction of false negatives, and f1-score is the product of the two (with a value of 1 for perfect classifiers). The low number of samples makes classifying into the four collisional outcomes challenging. Nevertheless, we find healthy metrics for hit-and-run and partial erosion classifications, with partial accretion being more challenging to classify.

\begin{table}
\centering
\begin{tabular}{ c c c c c} \hline 
Train CV & precision & recall & f1-score & samples \\
Classifier Data & & & & \\ \hline  
perfect accretion & 0.76 & 0.72 & 0.74 & 67 \\    
partial accretion & 0.81 & 0.84 & 0.82 & 110  \\
hit and run       & 0.93 & 0.92 & 0.93 & 155  \\
partial erosion   & 0.95 & 0.99 & 0.97 & 84 \\ \hline

Test Classifier Data     & precision & recall & f1-score & samples \\ \hline  
perfect accretion & 0.86 & 0.75 & 0.80 & 16 \\    
partial accretion & 0.67 & 0.73 & 0.70 & 11  \\
hit and run       & 0.81 & 0.87 & 0.84 & 15  \\
partial erosion   & 0.80 & 0.80 & 0.80 & 5 \\ \hline

Data for Regression    & precision & recall & f1-score & samples \\ \hline  
perfect accretion & 0.80 & 0.60 & 0.78 & 88 \\    
partial accretion & 0.71 & 0.81 & 0.77 & 108  \\
hit and run       & 0.92 & 0.96 & 0.92 & 178  \\
partial erosion   & 0.91 & 0.92 & 0.96 & 89 \\ \hline

\end{tabular} 
\caption{Table for Classification part of Nested models. Precision shows the percentage of false positives, recall, the percentage of false negatives and f1-score is the multiplication of both.}
\end{table}
\label{nested}

\begin{figure}
\begin{centering}
\includegraphics[width=\columnwidth]{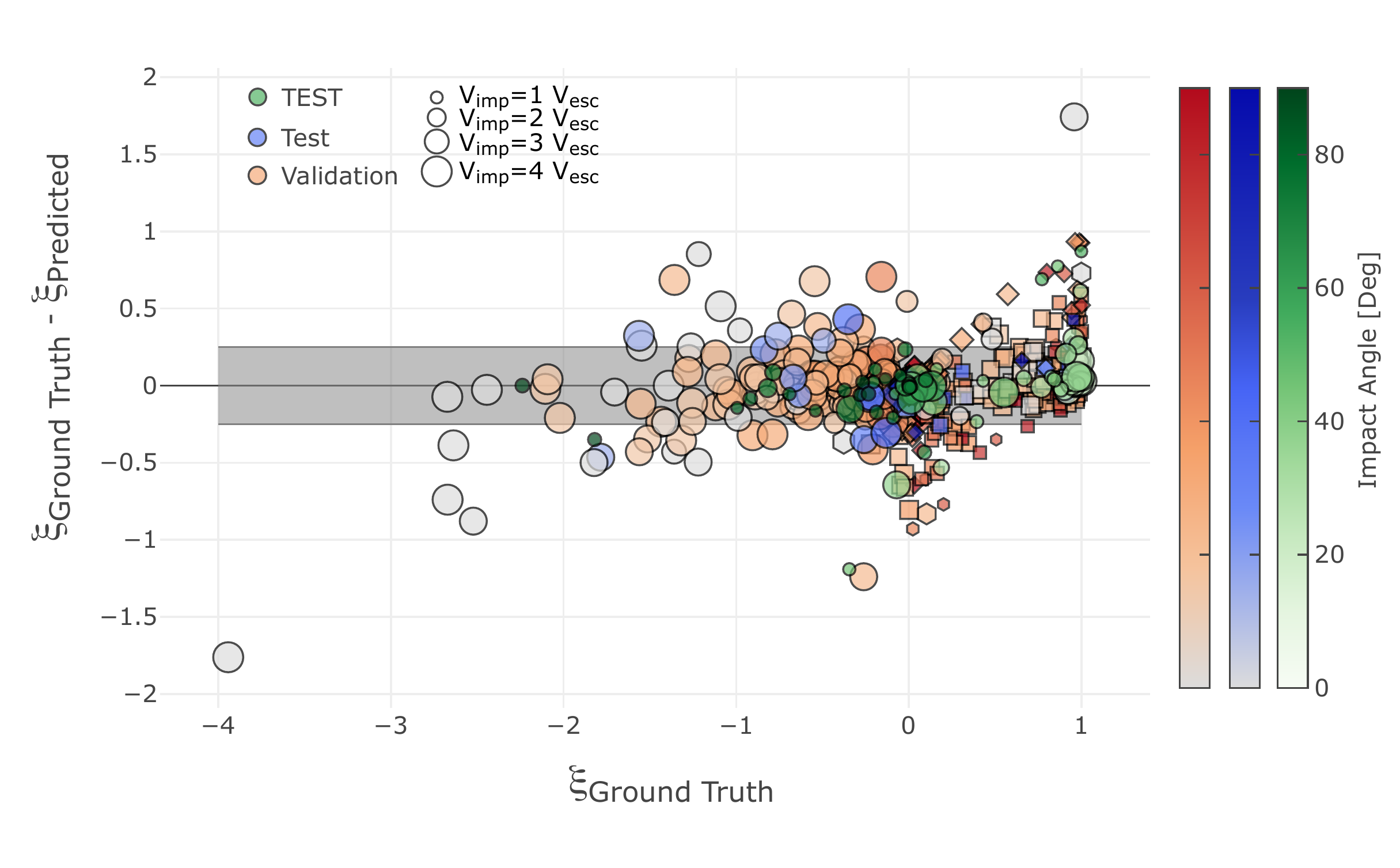}
\end{centering}

\caption{Differences between Ground Truth and Predicted values for Collisional Efficiency using Nested Models. Predictions on training set (red) and test set (blue) for each class (\emph{hexagon}: perfect accretion, \emph{squares:} partial accretion, \emph{diamond:}
hit-and-run, \emph{circle:} partial erosion) and then aggregated, as well as the overall test set data (green) are shown.   \label{fig:Nested}}

\end{figure}

After building the classification model, we fed it the data set aside and classified it into the four collisional regimes.  We subsequently built four regressor models with gradient boosting trees, one for each of the four types of collisions in the same way as described in section 4.1. The results are shown in Figure~\ref{fig:Nested}.
Another possibility is to implement the theoretical expectation for the collision types that are perfect accretion and hit-and-run instead of building a regressor. We experimented with setting $\xi = 0$ and $\xi=1$ for the data that was classified as perfect accretion or hit and run, but the method did slightly worse, probably because the threshold for labeling each class is somewhat arbitrary. 

We then used the features inspired by LS12 in the hopes it would help the classifier identify the collisional regime.  We did find an improvement in the classification of collisional outcomes, especially into partial accretion that reflected in an improvement on test data MSE (see Table 1). Using the geometrical engineered features provided similar gains.

\subsection{Gaussian Processes Regression}

One advantage of using gaussian processes is the ability to use the errors in the data as information, and obtain an error in the predictions. As the tools for processing SPH simulations calculate the mass of the largest remnant,  iteratively calculating which particles are gravitationally bound, there is an error associated with counting particles, and the mass they represent.  More importantly the outcomes of a giant impact include a considerable degree of stochasticity.  Running the same set of input parameters twice can result in slightly different results.  As such, we assumed a 0.05 and 0.1 absolute error on the efficiency. It may be that the error is not uniform (e.g. larger error for catastrophic or super catastrophic collisions), but as a first simple step, we consider it to be uniform. 

\begin{figure}
\begin{centering}
\includegraphics[width=\columnwidth]{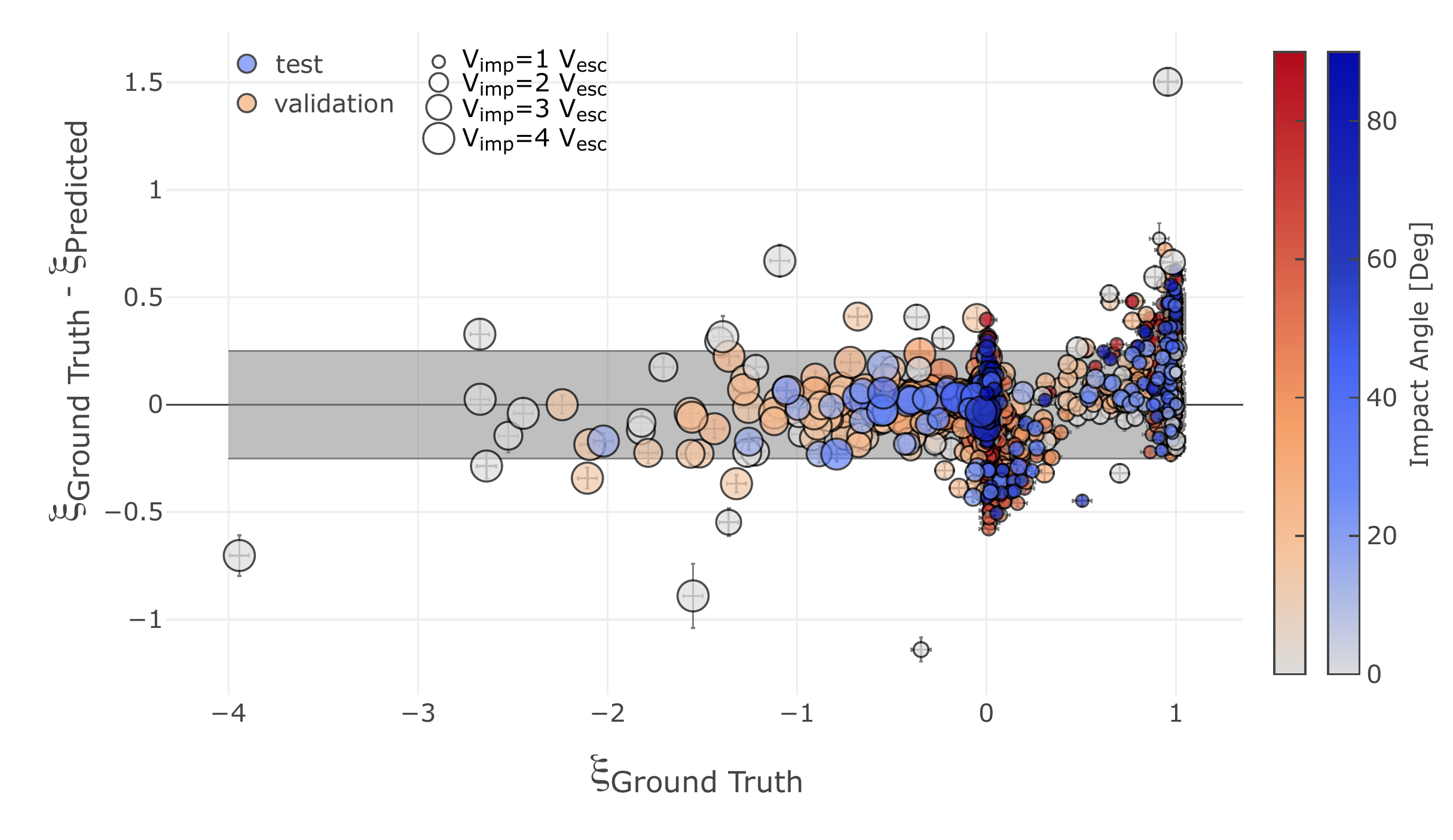}
\end{centering}

\caption{Differences between Ground Truth and Predicted values for Collisional Efficiency using Gaussian Processes.  Scheme same as Figure \ref{fig:Comp-GBTR-4F}. \label{fig:GP}}
\end{figure}

To ensure the scale of the data is the same (and for a more tractable fit of the hyper-parameters of the square exponentials), we perform a logarithmic transformation on the target mass first for the input features, and then scale-transform all features to the $[0,1]$ range. We use a training-test split of 80-20, and use hyper-opt to optimize the search and obtain the best values on the amplitude and length scale of the two square exponential kernels. As before, the results for the training with CV and test data are shown in Fig.\ref{fig:GP}.  It is important to note that all values above $\xi \ge 1$ are unphysical, but they are a consequence of a regression approach that does not have bounds.  

In sum, we found that decision trees outperformed our nested approach and gaussian processes, with the latter ones being at par. 

\subsection{Ensembling Models}

We examined the poorly-predicted samples at the 0.25 margin in all the models and found that there was considerable overlap. That is, there were 30-60 samples that were poorly-predicted in the validation or test sets by two or more models, and 19 samples that were poorly-predicted by all models. We then proceeded to investigate if ensembling could reduce the number of poorly-predicted samples and hence improve the overall performance. 

The idea behind ensembling models is to improve on the results that are poorly-predicted, while not changing the well-predicted samples. Often ensembling good models through averages produces better predictions. 

Thus, we ensembled the four models that improved the MSE on the test data through a simple average on the predictions. These models were:  gradient boosting trees with the four basic features, also with the geometrical features, also with the features from G19, as well as gaussian processes. We were not able to include the nested models in the ensemble because, by construction their test set is different. Only 23 samples coincided with the test sets of the other models, rendering it unviable for us to really assess prediction improvement. In addition, we found that using the model inspired by LS12 for ensembling did not improve the scores.   

\begin{figure*}
\begin{centering}
\includegraphics[width=\textwidth]{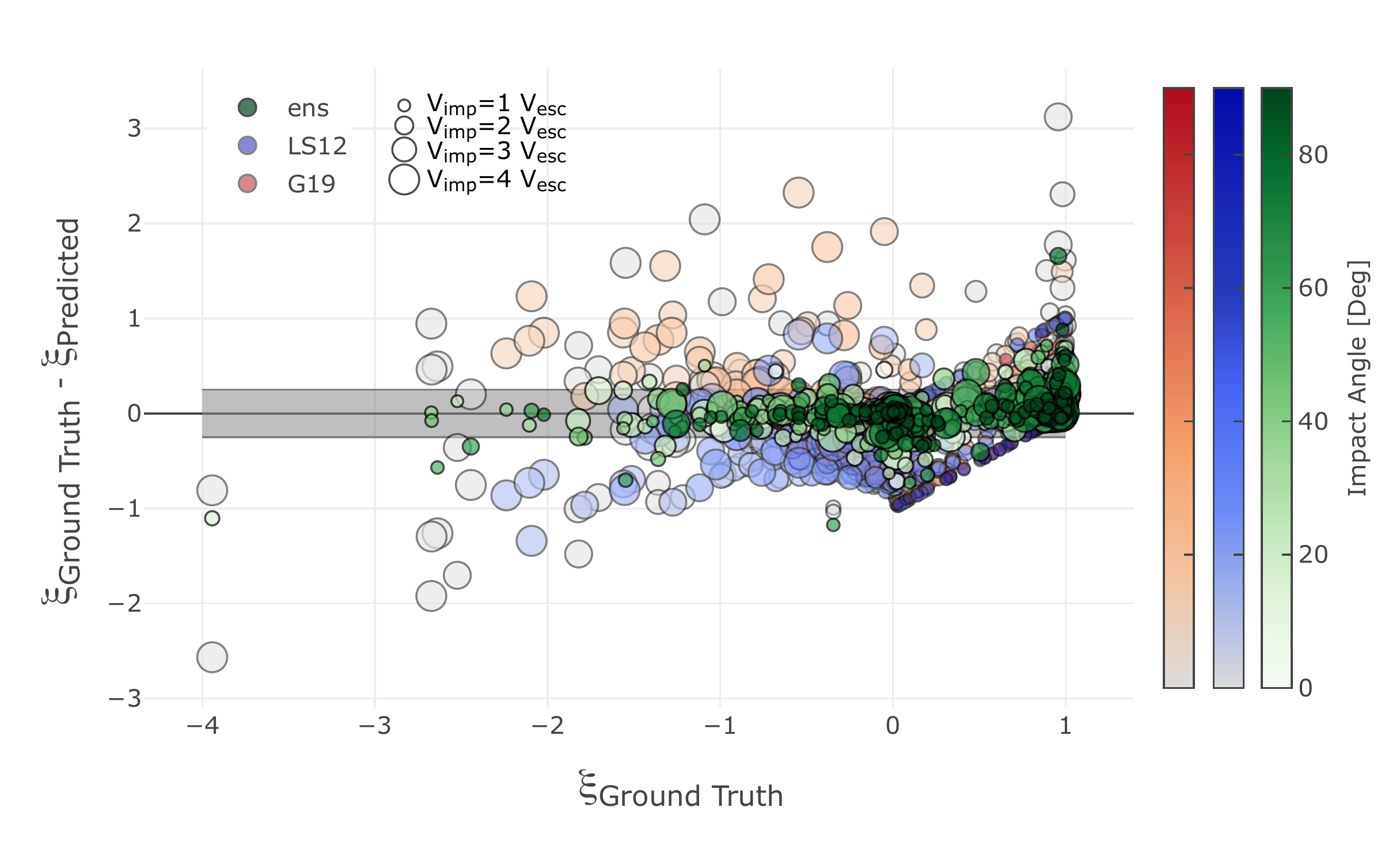}
\end{centering}

\caption{Differences between Ground Truth and Predicted values for Collisional Efficiency using our ensembled machine learning model (green). Predictions from LS12 (blue) and G19 (red)
are shown for reference. The 0.25 margin error is shown in grey.
 \label{fig:best}}
\end{figure*}

Our ensembling method improved the MSE to 0.0244 on the validation set, to 0.0373 on the test set and improved the number of samples well-predicted. This is our preferred model for predicting the mass of the largest remnant (via the efficiency) that yields an error of less than 0.25 margin in collisional efficiency for 92\% and less than 0.1 margin for 78\% of the predictions.
Figure \ref{fig:best} shows the overall prediction from this preferred model in $\xi$-space, compared to those inspired by the physical quantities proposed by LS12 and G19. Table 1 also shows overall metrics for LS12 and G19. Although it is not possible to compare exactly the performance among these models because the predictions on the validation set are usually overestimated,  it seems our machine learning strategies outperform the models by LS12 and G19.  

This final and all models are available upon request.

\section{Where to probe next}

\begin{figure*}
\begin{centering}
\includegraphics[width=\textwidth]{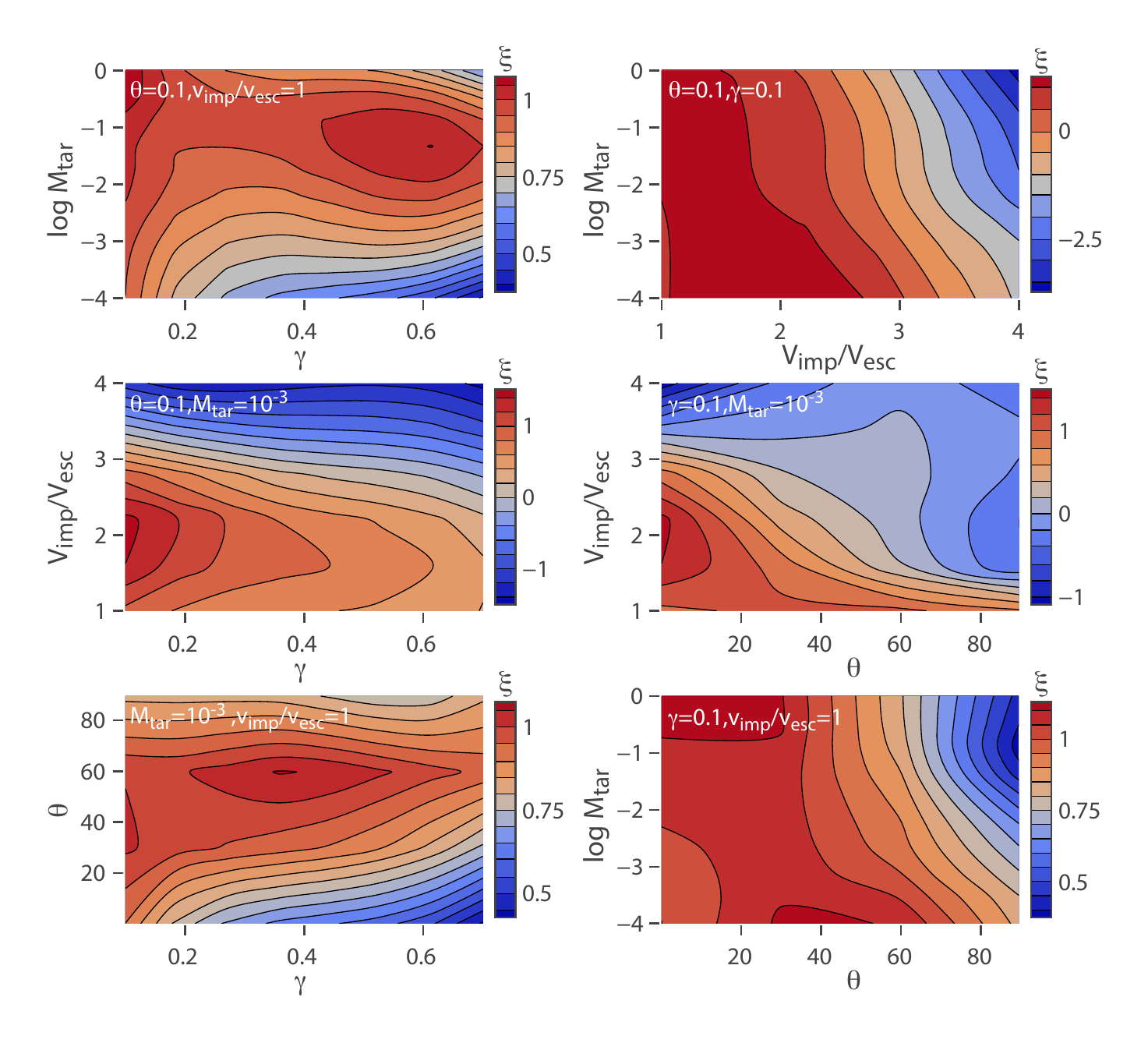}
\end{centering}

\caption{Predictions using the Gaussian-Processes-model on a grid of the input parameters target mass, mass ratio $\gamma$, impact angle $\theta$, and $v_{imp}/v_{esc}$ shown as 3D cuts by keeping constant two input parameters at a time (shown in each figure). 
\label{fig:grid_GP}}
\end{figure*}

\begin{figure*}
\begin{centering}
\includegraphics[width=\textwidth]{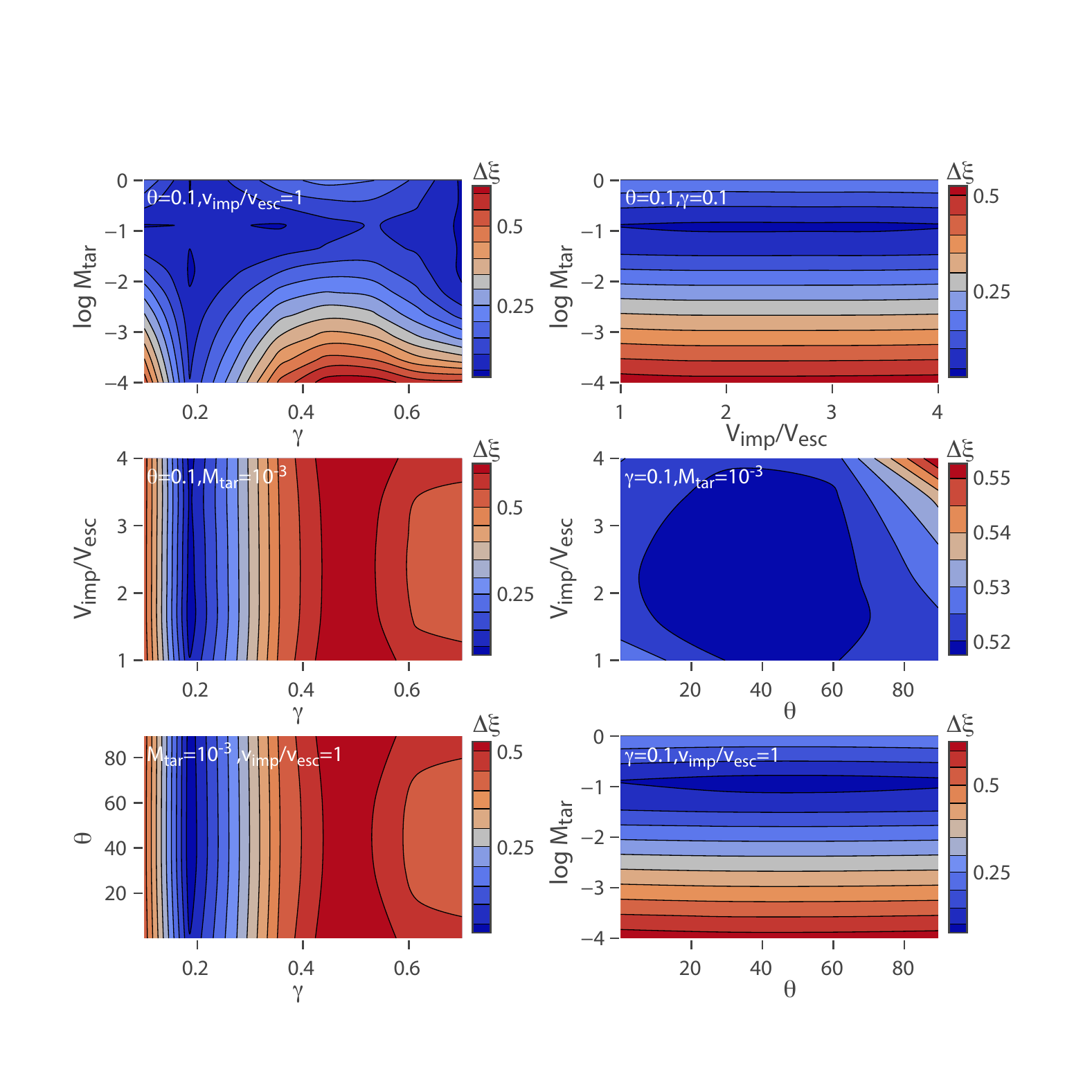}
\end{centering}

\caption{Errors associated with the collisional efficiency predictions shown in Fig. \ref{fig:grid_GP} .
\label{fig:griderror_GP}}

\end{figure*}

The exercise of modelling a collision with SPH simulations is time consuming, and the parameter space to explore is vast. Thus, it is key to understand what are the most significant parameters to explore first to enhance the predictions of machine learning algorithms. As a first approach, we used the power of gaussian processes in determining an error in the predictions and searched for where this error was largest.  We predicted values on a grid that spanned the ranges of the original data: the four orders of magnitude in target mass, mass ratios from 0.1 to 0.7, impact velocities of 1 to 4 times the escape velocity, and impact angles from 0.1 to 89.5 degrees. The result is a 5 dimensional function with 4 basic problem parameters and 1 prediction for the collisional efficiency.  This 5D-structure is difficult to visualize, thus, we selected a few cuts to illustrate in 3D some of its properties. We show the predicted collisional efficiency in Fig. \ref{fig:grid_GP} and associated errors  in Fig. \ref{fig:griderror_GP}.

It is clear that the impact angle and $v_{\rm imp}/v_{\rm esc}$ are the most well-sampled parameters in the collisional data. In contrast, there is a well established maximum in the errors associated with $\gamma\simeq 0.5$ regardless of angle, or impact velocity. Low values for target mass ($1\times10^{-3}M_E$) also seem under-sampled. 

\begin{figure}
\begin{centering}
\includegraphics[width=\columnwidth]{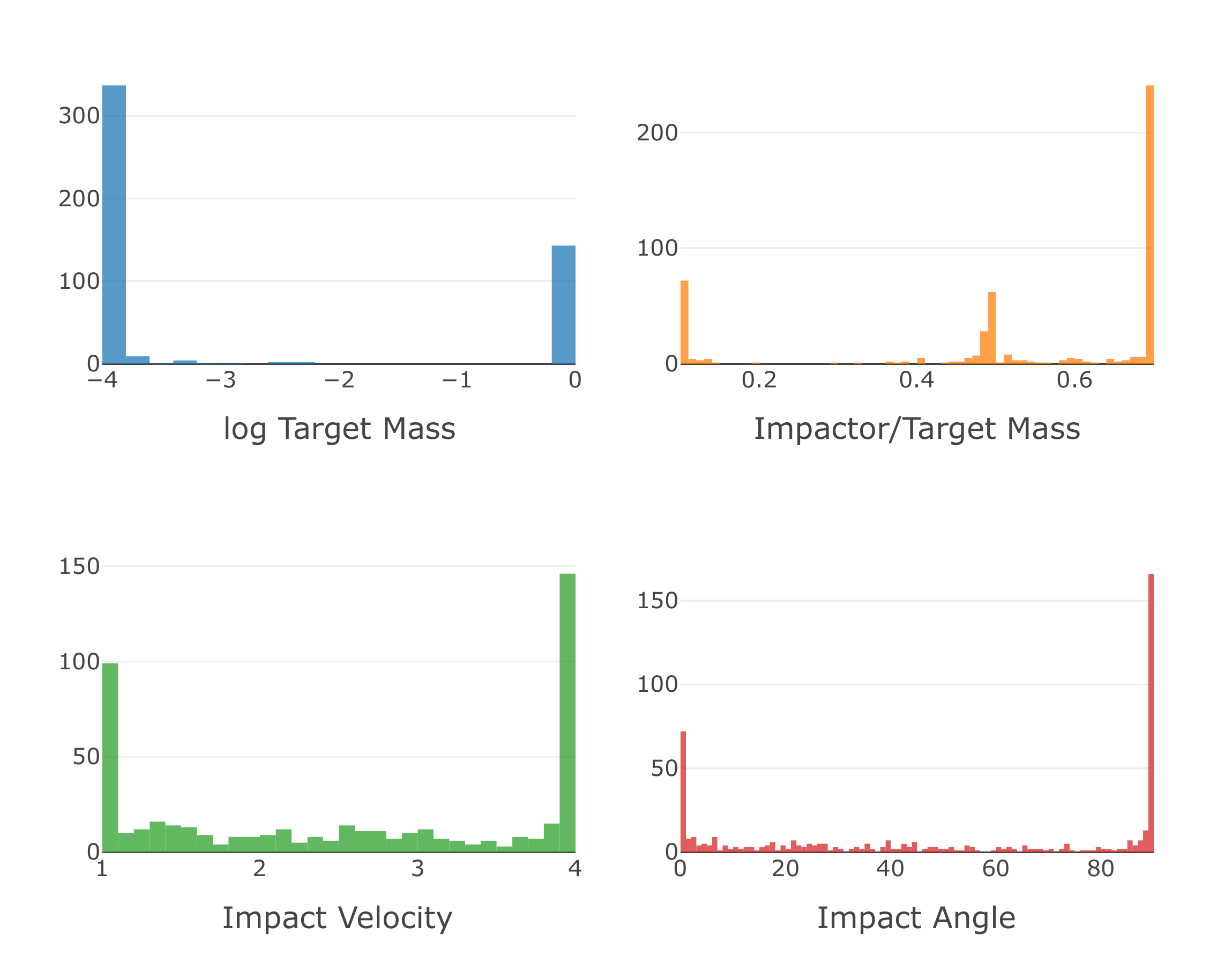}
\end{centering}

\caption{Histograms of input parameters that lead to maxima of the prediction-errors function obtained with Gaussian Processes. \label{fig:Histograms-maxima}}

\end{figure}

To obtain more quantitative estimates of where the maxima of the prediction-errors function are and go beyond a visual inspection, we searched for the locations where the predicted errors were largest. We applied a maximization scheme on the predicted error by picking an initial guess that we drew randomly (random draw of the four basic features). We did this 500 times, and obtained multiple maxima which we organized into a histogram of each of the four input parameters. Figure \ref{fig:Histograms-maxima} shows the number of times each value of the basic features led to a maxima in the prediction errors.  In other words, the histograms show which values for input parameters consistently yielded large uncertainties in predictions.

Based on these results, our suggestion is to run SPH collision calculations that preferentially sample target masses of $10^{-4}$ and 1~$M_{\oplus}$, mass ratios of $0.5$ and $0.7$ and impact velocities of 1~$v_{\rm esc}$. Values outside the range of the data explored are expected to have even larger errors in the prediction (e.g. mass ratio $>0.7$).  Note that while the errors are also larger at $\left(v_{\rm imp}/v_{\rm esc}\right)\sim 4$ collisions at such high velocities are very rare (see G19) and thus additional simulations to reduce the errors in this region are not as valuable.

\section{Summary}

In summary, we used supervised machine learning algorithms on collisional data, where the objective was to predict the mass of the largest remnant via the efficiency factor $\left(\xi=\frac{M_{\rm LR}-M_{\rm tar}}{M_{\rm imp}}\right)$.  The three models tested were gradient boosting regression trees, nested models (classification into four collisional regimes followed by regression in each class), and gaussian processes. The best results were produced by the gradient boosting regression trees, with a minor improvement by ensembling different algorithms with a standard average. Our best model was able to predict 88\% within a 0.25 confidence margin with unseen data.  It performs better than physically-inspired existing collisional outcome maps. 

We also tried adding to the four basic features, geometrical features that aimed at separating the gentle (low impact velocity and angle) from the violent collisions (high impact velocity and low impact angle) with slight improvement.

Finally, we tested engineered features based on two predictions inspired by physical models. Both of these models have a prescription based on the ratio of impact energies to some critical disruption energy, and a fit to their data. They also have switches that aim to constrain the different types of collisions. We found that there was no gain in using these physically-inspired features to the model. Thus, concluding that the four basic features of target mass, mass ratio, impact to escape velocity, and impact angle are sufficient for the model to make predictions.

By using the power of gaussian processes in predicting the variance associated with each prediction, we ran a grid of input parameters to determine what parameter values would provide the most gain in information to machine learning algorithms given the current data.  The most prominent feature is mass ratios above 0.5, followed by target masses of $\sim10^{-4}M_{Earth}$, as well as the lowest impact velocities of $\sim 1 v_{esc}$.


This study is one of the first to implement a supervised machine learning approach to predictions of collisional outcomes. The fact that with only a modest data set (1029 samples) we were able to make meaningful predictions on the mass of the largest remnant, it may be indicative that by growing the data set the machine can learn collisional outcomes in a way that is useful to formation models. This may be the beginning of a new direction of collisional studies. 

\acknowledgements{}

DV is supported by the Natural Sciences and Engineering Research Council of Canada ( grant RBPIN-2014-06567). APJ is supported by the Centre for Planetary Sciences at the University of Toronto, Scarborough.We would like to acknowledge that our work was performed on land traditionally inhabited by the Wendat, the Anishnaabeg, Haudenosaunee, Metis and the Mississaugas of the New Credit First Nation.

\bibliographystyle{aasjournal}
\bibliography{references}

\begin{thebibliography}{}
\expandafter\ifx\csname natexlab\endcsname\relax\def\natexlab#1{#1}\fi
\providecommand{\url}[1]{\href{#1}{#1}}
\providecommand{\dodoi}[1]{doi:~\href{http://doi.org/#1}{\nolinkurl{#1}}}
\providecommand{\doeprint}[1]{\href{http://ascl.net/#1}{\nolinkurl{http://ascl.net/#1}}}
\providecommand{\doarXiv}[1]{\href{https://arxiv.org/abs/#1}{\nolinkurl{https://arxiv.org/abs/#1}}}

\bibitem[{{Agnor} \& {Asphaug}(2004)}]{Agnor_Asphaug:2004}
{Agnor}, C., \& {Asphaug}, E. 2004, ApJL, 613, L157, \dodoi{10.1086/425158}

\bibitem[{{Ambikasaran} {et~al.}(2014){Ambikasaran}, {Foreman-Mackey},
  {Greengard}, {Hogg}, \& {O'Neil}}]{george}
{Ambikasaran}, S., {Foreman-Mackey}, D., {Greengard}, L., {Hogg}, D.~W., \&
  {O'Neil}, M. 2014

\bibitem[{{Asphaug}(2010)}]{asphaug2010}
{Asphaug}, E. 2010, Chemie der Erde / Geochemistry, 70, 199,
  \dodoi{10.1016/j.chemer.2010.01.004}

\bibitem[{{Asphaug} \& {Reufer}(2014)}]{asphaug2014}
{Asphaug}, E., \& {Reufer}, A. 2014, Nature Geoscience, 7, 564,
  \dodoi{10.1038/ngeo2189}

\bibitem[{{Bergstra} {et~al.}(2015){Bergstra}, {Komer}, {Eliasmith}, {Yamins},
  \& {Cox}}]{hyperopt}
{Bergstra}, J., {Komer}, B., {Eliasmith}, C., {Yamins}, D., \& {Cox}, D.~D.
  2015, Computational Science and Discovery, 8, 014008,
  \dodoi{10.1088/1749-4699/8/1/014008}

\bibitem[{{Bond} {et~al.}(2010){Bond}, {O'Brien}, \& {Lauretta}}]{Bond:2010}
{Bond}, J.~C., {O'Brien}, D.~P., \& {Lauretta}, D.~S. 2010, ApJ, 715, 1050,
  \dodoi{10.1088/0004-637X/715/2/1050}

\bibitem[{{Canup} \& {Asphaug}(2001)}]{canup2001}
{Canup}, R.~M., \& {Asphaug}, E. 2001, Nature, 412, 708,
  \dodoi{10.1038/35089010}

\bibitem[{{Chambers}(2001)}]{Chambers:2001}
{Chambers}, J.~E. 2001, Icarus, 152, 205, \dodoi{10.1006/icar.2001.6639}

\bibitem[{{Chambers}(2013)}]{Chambers:2013}
---. 2013, Icarus, 224, 43, \dodoi{10.1016/j.icarus.2013.02.015}

\bibitem[{{Chen} \& {Guestrin}(2016)}]{Xgboost}
{Chen}, T., \& {Guestrin}, C. 2016, ArXiv e-prints.
\newblock \doarXiv{1603.02754}

\bibitem[{{Genda} {et~al.}(2011){Genda}, {Kokubo}, \& {Ida}}]{genda2011}
{Genda}, H., {Kokubo}, E., \& {Ida}, S. 2011, in Lunar and Planetary Science
  Conference, Vol.~42, Lunar and Planetary Science Conference, 2090

\bibitem[{{Kenyon} \& {Bromley}(2006)}]{kenyon2006}
{Kenyon}, S.~J., \& {Bromley}, B.~C. 2006, Astronomical Journal, 131, 1837,
  \dodoi{10.1086/499807}

\bibitem[{{Kokubo} \& {Genda}(2010)}]{kokubo2010}
{Kokubo}, E., \& {Genda}, H. 2010, Astrophysical Journal Letters, 714, L21,
  \dodoi{10.1088/2041-8205/714/1/L21}

\bibitem[{{Kokubo} \& {Ida}(2002)}]{Kokubo:2002}
{Kokubo}, E., \& {Ida}, S. 2002, ApJ, 581, 666, \dodoi{10.1086/344105}

\bibitem[{{Leinhardt} \& {Stewart}(2012)}]{Leinhardt_Stewart:2012}
{Leinhardt}, Z.~M., \& {Stewart}, S.~T. 2012, Apj, 745, 79,
  \dodoi{10.1088/0004-637X/745/1/79}

\bibitem[{{Marcus} {et~al.}(2010{\natexlab{a}}){Marcus}, {Sasselov},
  {Hernquist}, \& {Stewart}}]{marcus2010a}
{Marcus}, R.~A., {Sasselov}, D., {Hernquist}, L., \& {Stewart}, S.~T.
  2010{\natexlab{a}}, \apjl, 712, L73, \dodoi{10.1088/2041-8205/712/1/L73}

\bibitem[{{Marcus} {et~al.}(2010{\natexlab{b}}){Marcus}, {Sasselov}, {Stewart},
  \& {Hernquist}}]{marcus2010b}
{Marcus}, R.~A., {Sasselov}, D., {Stewart}, S.~T., \& {Hernquist}, L.
  2010{\natexlab{b}}, \apjl, 719, L45, \dodoi{10.1088/2041-8205/719/1/L45}

\bibitem[{{Marcus} {et~al.}(2009){Marcus}, {Stewart}, {Sasselov}, \&
  {Hernquist}}]{marcus2009}
{Marcus}, R.~A., {Stewart}, S.~T., {Sasselov}, D., \& {Hernquist}, L. 2009,
  Astrophysical Journal Letters, 700, L118,
  \dodoi{10.1088/0004-637X/700/2/L118}

\bibitem[{{Melosh}(2007)}]{melosh2007}
{Melosh}, H.~J. 2007, Meteoritics and Planetary Science, 42, 2079,
  \dodoi{10.1111/j.1945-5100.2007.tb01009.x}

\bibitem[{{Movshovitz} {et~al.}(2016){Movshovitz}, {Nimmo}, {Korycansky},
  {Asphaug}, \& {Owen}}]{movshovitz2016}
{Movshovitz}, N., {Nimmo}, F., {Korycansky}, D.~G., {Asphaug}, E., \& {Owen},
  J.~M. 2016, \icarus, 275, 85, \dodoi{10.1016/j.icarus.2016.04.018}

\bibitem[{{Mustill} {et~al.}(2018){Mustill}, {Davies}, \&
  {Johansen}}]{Mustill:2018}
{Mustill}, A.~J., {Davies}, M.~B., \& {Johansen}, A. 2018, MNRAS, 478, 2896,
  \dodoi{10.1093/mnras/sty1273}

\bibitem[{{Rasmussen} \& {Williams}(2006)}]{GPbook}
{Rasmussen}, C.~E., \& {Williams}, C.~K.~I. 2006, {Gaussian Processes for
  Machine Learning}

\bibitem[{{Raymond} {et~al.}(2009){Raymond}, {O'Brien}, {Morbidelli}, \&
  {Kaib}}]{raymond2009}
{Raymond}, S.~N., {O'Brien}, D.~P., {Morbidelli}, A., \& {Kaib}, N.~A. 2009,
  Icarus, 203, 644, \dodoi{10.1016/j.icarus.2009.05.016}

\bibitem[{{Raymond} {et~al.}(2004){Raymond}, {Quinn}, \&
  {Lunine}}]{Raymond:2004}
{Raymond}, S.~N., {Quinn}, T., \& {Lunine}, J.~I. 2004, Icarus, 168, 1,
  \dodoi{10.1016/j.icarus.2003.11.019}

\bibitem[{{Terquem} \& {Papaloizou}(2007)}]{Terquem:2007}
{Terquem}, C., \& {Papaloizou}, J.~C.~B. 2007, ApJ, 654, 1110,
  \dodoi{10.1086/509497}

\end{thebibliography}

\end{document}